\documentclass[acmtog]{acmart}
\AtBeginDocument{%
  \providecommand\BibTeX{{%
    \normalfont B\kern-0.5em{\scshape i\kern-0.25em b}\kern-0.8em\TeX}}}

\copyrightyear{2024}
\acmYear{2024}
\setcopyright{acmlicensed}\acmConference[SA Conference Papers '24]{SIGGRAPH Asia 2024 Conference Papers}{December 3--6, 2024}{Tokyo, Japan}
\acmBooktitle{SIGGRAPH Asia 2024 Conference Papers (SA Conference Papers '24), December 3--6, 2024, Tokyo, Japan}
\acmDOI{10.1145/3680528.3687624}
\acmISBN{979-8-4007-1131-2/24/12}

\acmSubmissionID{566}

\usepackage{cleveref}
\usepackage{subcaption}
\usepackage{wrapfig}
\usepackage{etoolbox}
\usepackage[normalem]{ulem}
\makeatletter
\patchcmd\WF@putfigmaybe{\lower\intextsep}{}{}{\fail}%
\AddToHook{env/wrapfigure/begin}{\setlength{\intextsep}{0pt} \setlength{\columnsep}{0pt}}
\makeatother

\usepackage{xcolor}

\newcommand{\ourmethod}[0]{\textit{Casper DPM}}
\begin{document}

\title[\ourmethod]{\ourmethod:\\ Cascaded Perceptual Dynamic Projection Mapping onto Hands}

\author{Yotam Erel}
\email{erelyotam@gmail.com}
\orcid{0000-0001-8319-5111}
\affiliation{%
  \institution{Tel Aviv University}
  \city{Tel Aviv}
  \country{Israel}
}

\author{Or Kozlovsky-Mordenfeld}
\affiliation{%
  \institution{Tel Aviv University}
  \city{Tel Aviv}
  \country{Israel}
}

\author{Daisuke Iwai}
\affiliation{%
  \institution{Osaka University}
  \city{Osaka}
  \country{Japan}
}

\author{Kosuke Sato}
\affiliation{%
  \institution{Osaka University}
  \city{Osaka}
  \country{Japan}
}

\author{Amit H. Bermano}
\affiliation{%
 \institution{Tel Aviv University}
  \city{Tel Aviv}
  \country{Israel}
}


\begin{abstract}
  We present a technique for dynamically projecting 3D content onto human hands with short perceived motion-to-photon latency. Computing the pose and shape of human hands accurately and quickly is a challenging task due to their articulated and deformable nature. We combine a slower 3D coarse estimation of the hand pose with high speed 2D correction steps which improve the alignment of the projection to the hands, increase the projected surface area, and reduce perceived latency. Since our approach leverages a full 3D reconstruction of the hands, any arbitrary texture or reasonably performant effect can be applied, which was not possible before. We conducted two user studies to assess the benefits of using our method. The results show subjects are less sensitive to latency artifacts and perform faster and with more ease a given associated task over the na\"ive approach of directly projecting rendered frames from the 3D pose estimation. We demonstrate several novel use cases and applications.
\end{abstract}



\begin{teaserfigure}
\centering
  \includegraphics[width=\textwidth]{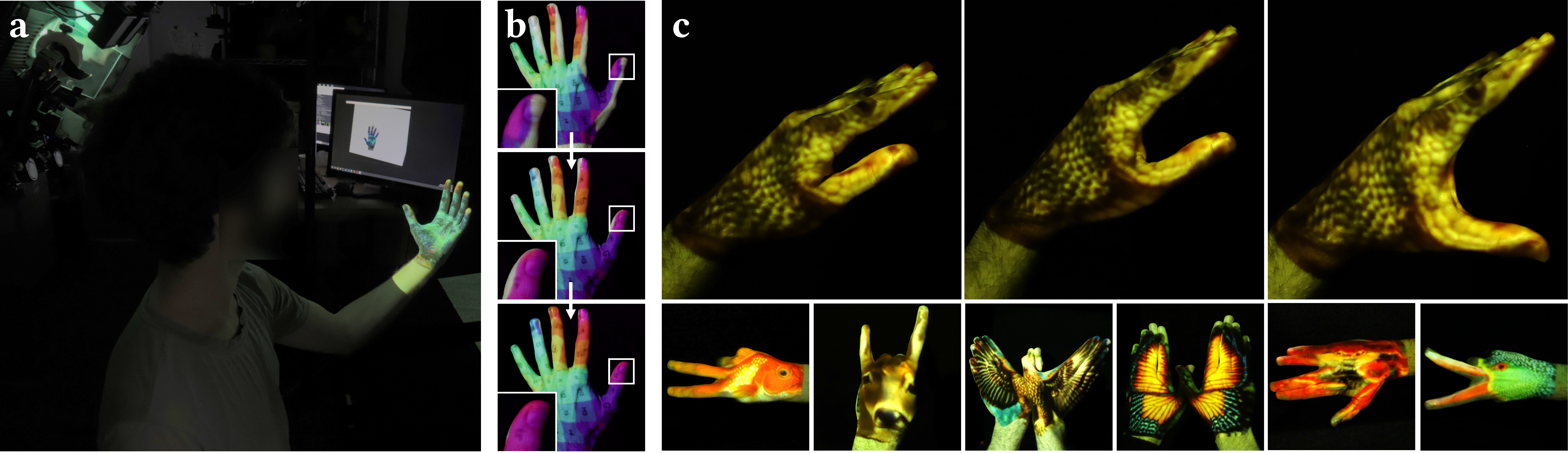}
  \caption{\ourmethod~employs a single projector to augment the appearance of human hands in real-time (a). Our cascaded perceptual pipeline corrects 3D tracking misalignment (b top) by 2D deformations (b middle), and then quickly reducing the boundaries formed between the projection and the hand (b bottom). Content can be automatically generated to fit the hand pose, or preloaded as UV texture (c bottom). Real time hand and finger performance drives the content accordingly, allowing perceptually pleasing augmentations (c top).}
  \Description{A dynamic projection mapping is applied to two human hands in real time.}
  \label{fig:teaser}
\end{teaserfigure}


\maketitle

\definecolor{regb}{RGB}{191, 44, 35}
\definecolor{regc}{RGB}{47, 103, 177}
\definecolor{regd}{RGB}{64, 176, 166}

\section{Introduction}
\label{sec:intro}
Dynamic Projection Mapping (DPM) is a highly challenging task yet with numerous visually pleasing applications. Augmenting the appearance of moving objects can be both entertaining and informative, e.g. as employed by amusements parks \cite{amusement_dpm}, product design \cite{moji_fabric_flex}, makeup \cite{makeuplamps}, collaboration support \cite{omnilantern}, medicine \cite{medical_dpm} and more. However, dynamic projector based augmentations requires high precision and even higher performance. Humans are sensitive to even slight misalignment during projection \cite{6ms_latency, user_misalignment} --- a trait very difficult to satisfy for moving objects, as they need to be tracked in 3D, processed to projection images, and passed to the projector, while they continue to move. 

This is especially true for human hands, which are highly articulate, and tend to move rapidly and sporadically. For this reason, even though hands are our most effective manipulation tools as humans, DPM onto body parts techniques in the literature address mostly facial and full-body motion, but rarely projection onto human hands. Related methods that do address hands impose restrictions to relax the problem (e.g., using wearables \cite{ring_imu}, using markers or avoid the palm area \cite{arm_dpm}, accepting lower latencies \cite{old_pm_hands3}, restricting the type of augmentations possible \cite{midas}, etc.).

In this paper we introduce \ourmethod~ --- a short motion-to-photon projection mapping framework that enables wearable-free full 3D augmentations with low perceived latency. Our cascaded solution combines a coarse pose estimator in 3D with accurate and rapid screen space corrections based on a high speed camera. With it, we first correct low frequency artifacts, including calibration misalignment, shape mismatches, skinning and tracking errors. Still, after this step edges do not align well during performance due to shape mismatches and latency issues. Hence, to improve perceived latency, we further exploit the attention given by our perception of motion to silhouette edges, and propose a last-minute correction to cover the whole observed surface area. 

The interplay between the different asynchronous steps causes various temporal artifacts, difficult to balance. We hence also propose a discrete time simulation of our system, which we used to analyze different visual artifacts, identify their root cause, and propose a balanced way to reduce them. 

Equipped with these insights, we have implemented a prototype for evaluations, and have developed different experimental and entertaining applications. Our solution is the first to tackle DPM onto wearable-free hands with full 3D capabilities. Related studies are hardware specific or otherwise irreproducible, and do not present quantitative metrics for comparisons. Therefore we conduct experiments comparing our method to the standard track-render-project approach. Through two user studies, one in simulation and one live, we show that \ourmethod~offers more immersive results, where viewers are less disturbed by misalignment artifacts, and perform better at tasks. 

We envision our system enabling simple and immersive DPM onto hands, allowing new applications such as gaming, physiotherapy, social performances, and creative artwork.

In summary, the core contributions of this paper are:
\begin{itemize}
    \item A system allowing for a highly immersive experience of dynamically projecting content onto human hands in real-time.
    \item A method to correct 3D reconstruction of human hands using 2D projections of the same underlying shape, enabling better precision in coverage, especially of the fingers.
    \item A novel 2D perceptual boundary reduction step that reduces the perceived notion of latency by exploiting human attention given to silhouette edges.
\end{itemize}

An open sourced implementation of the graphics engine driving our system, a discrete time simulator, described applications and experimental results will be made publicly available.

\section {Related Work}
\label{sec:related}

Early attempts to perform projection mapping onto human hands began to appear with the emergence of new depth sensors and advances in classical computer vision techniques \cite{old_pm_hands0, old_pm_hands1, old_pm_hands2, old_pm_hands3}. In more recent studies, DPM with short latencies and high throughput has been successfully applied to a variety of objects including both rigid inanimate objects and non-rigid human body parts \cite{dpm_rigid, makeuplamps, arm_dpm, midas, faceforge, dpm_sil, dpm_performer_sil}. Achieving short latencies is essential for creating highly immersive experiences and is made possible through the combination of high-speed hardware (cameras, projectors, lighting) and robust algorithms (tracking, rendering). In recent years, there have been tremendous improvements in the quality and speed of these elements, facilitating the advancements of applying DPM to new domains \cite{pm_survey}. Interestingly, high quality DPM onto human hands was rather avoided so far by previous studies despite their appealing applications. We speculate this is because tracking and reconstructing them is a significantly harder challenge than other human body parts in terms of accuracy, robustness and speed.

One way to alleviate these problems is to use low-latency wearable sensors such as gloves \cite{glove_imu} or rings \cite{ring_imu}. However, reconstructing a wearable-free hand pose and shape robustly in a few milliseconds, even from multiple cameras, is still an out of reach goal despite numerous recent attempts to push the bar \cite{mediapipe, mobrecon, quest2_handtracking, quest3_handtracking}. In VR, where alignment between the virtual asset and real hand is not an issue, these tracking solutions are good enough - but misalignments are noticeable for video see-through displays such as in commercial headsets (e.g. Meta Quest 3, Apple Vision Pro), and certainly more disturbing when performing projection mapping where any slight misalignment will cause a noticeable difference \cite{user_misalignment}. Projection mapping imposes another unique constraint on tracking, which is the projection illumination interfering with acquisition, hence the need for some sort of band separation, e.g. by restricting the acquisition to the IR spectrum, which may reduce the quality of visible-light based solutions. 

\citet{arm_dpm} proposed a system to project onto human arms considering skin deformations with barely any perceivable latency by modeling the arm as a parametric deformable surface. However, this is done by using passive markers, which are less suitable for hands, as many such markers need to be prepared in advance per user, hindering the flow and efficiency of the downstream application significantly. \citet{dpm_deformables} use B-Spline patches to model surface deformations and project dynamic content onto them, but this limits their targets to stretchable cloths, and is not suitable for projection onto highly articulated and self-occluding objects like human hands. In MIDAS \cite{midas}, the authors propose a method to augment any object, including deformables, by rapidly reconstructing its normals using a shape from shading technique and three band-separated high speed cameras. While the results are impressive, this method does not allow for general spatially varying augmentations to be performed other than tileable textures. Our system on the other hand does not impose such constraint, as the full 3D hand pose is estimated for skinning a mesh. 

It is fairly clear that in the current state of technology, tracking and reconstructing (wearable-free) hands accurately with very short latencies is infeasible. Thus, instead of tackling these core issues, we embrace a more ad-hoc approach and devise a cascaded solution based on a coarse 3D sensor and rapid correction steps in 2D, tailored specifically for projection mapping onto hands and how it is perceived by a user.

\section{Method}

\begin{figure*}[t]
  \centering
  \includegraphics[width=0.9\hsize]{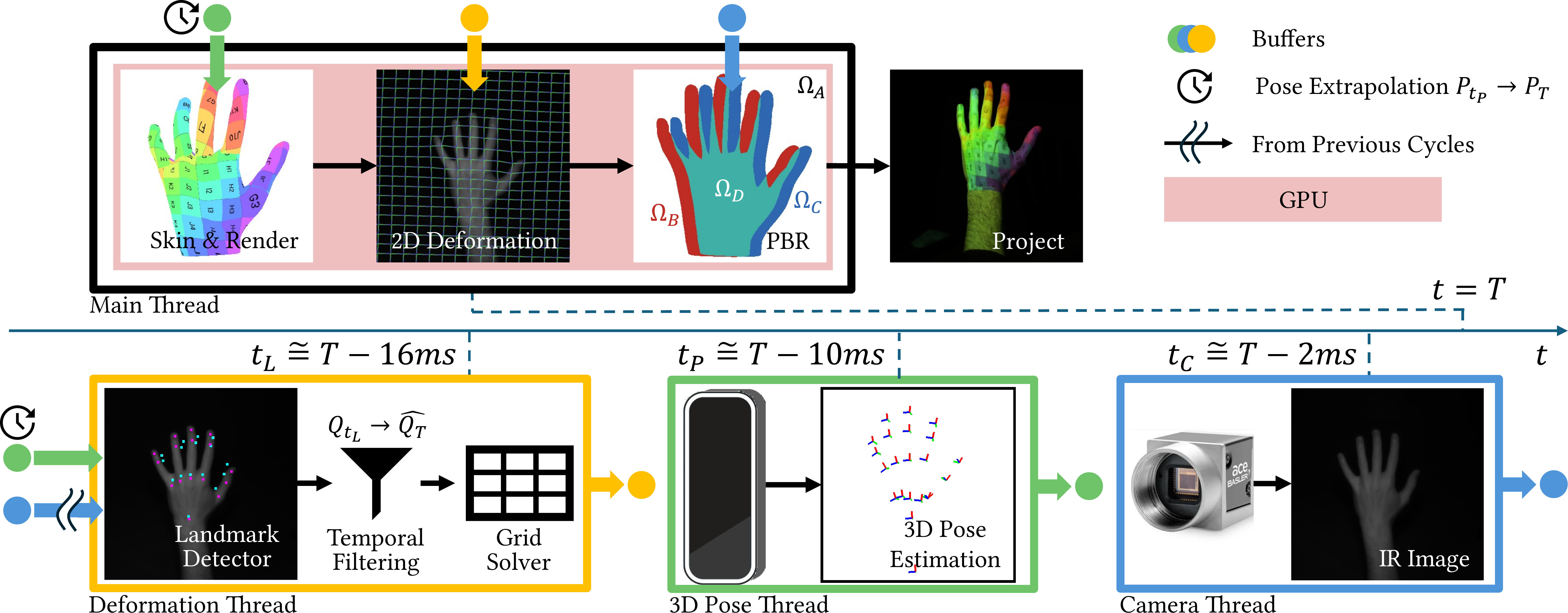}
  \caption{Pipeline overview. Top: for every projected frame (at time $T$) our main thread generates and skins a 3D mesh using an extrapolated 3D pose estimation (\Cref{sec:motion_to_photon}), deforms the rendered mesh in 2D to reduce static bias using 2D landmark motion estimation (\Cref{sec:mls_deformation}), and performs a novel perceptual boundary reduction step (\Cref{sec:pbr}) which decreases the perceived latency,  extending the projected surface area to match precise hand position rapidly. The result is projected onto the hand. Bottom: three asynchronous threads are responsible for filling the buffers with information, each operating asynchronously at different rates.}
  \label{fig:pipeline}
  \Description{A block diagram showing the whole system decomposed into discrete operations. The top shows blocks operating at the current time frame, and the bottom shows what happens asynchronously before the current time frame. Top and bottom are separated by a timeline axis. There is a legend on the top right indicating the type of connections between the blocks: buffers, synchronous connections, connections from previous cycles and asynchronous connections.}
\end{figure*}

\begin{figure}[h]
  \centering
  \includegraphics[width=0.9\hsize]{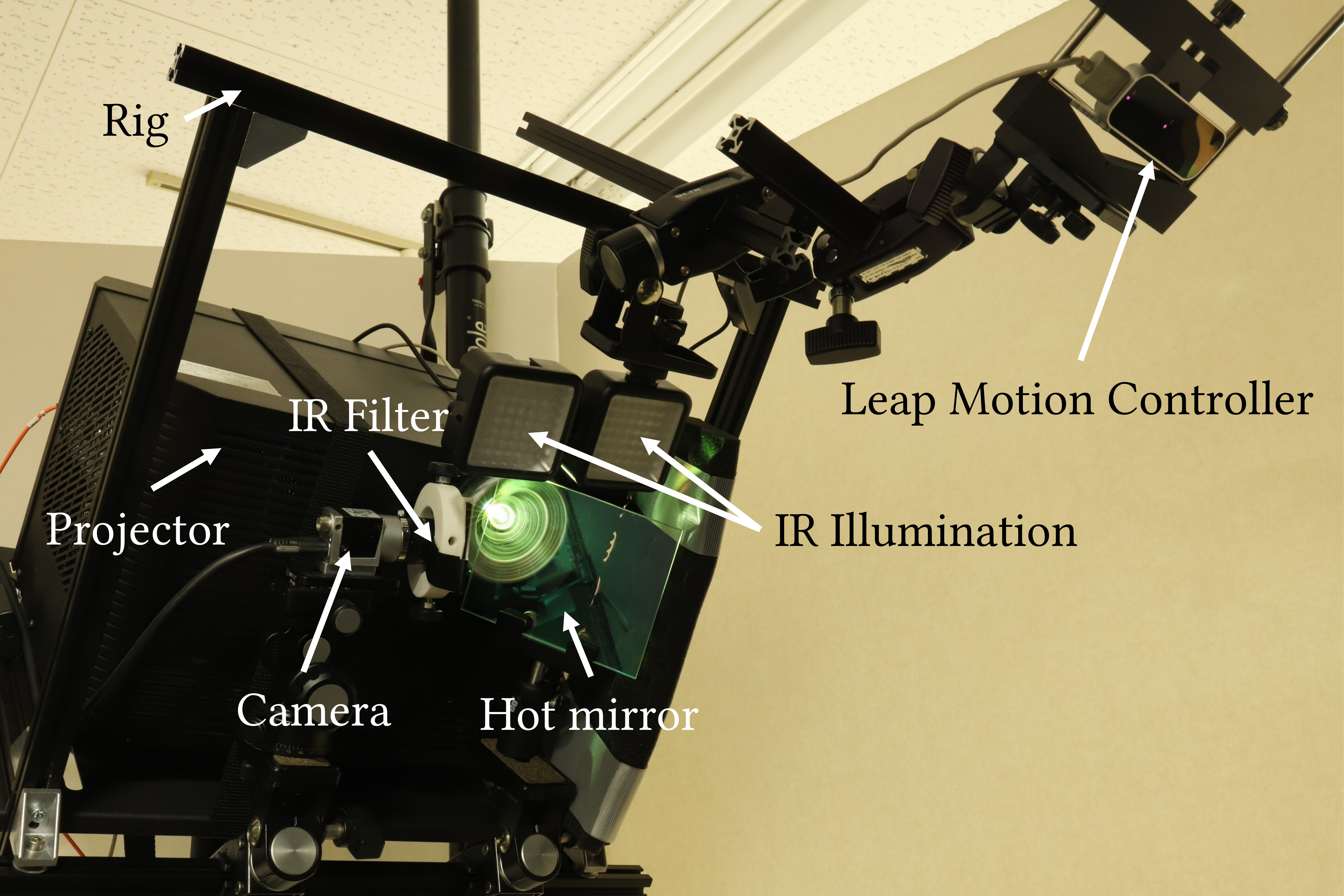}
  \caption{Physical setup. The setup consists of a coaxial high speed projector and camera, diffuse IR illumination, a hot mirror, and a Leap Motion Controller, held together on a configurable rig.}
  \Description{A photo showing the physical elements of the system.}
  \label{fig:physical_setup}
\end{figure}

\label{sec:method}
A schematic overview of \ourmethod~is shown in \Cref{fig:pipeline}, and the physical setup in \Cref{fig:physical_setup}. 
In terms of hardware, we employ a typical projection-camera setup \cite{moji_fabric_flex, moji_proj_nonrig, makeuplamps, context-aware-light-source}, where a high-speed projector operates in the visible spectra, and a coaxial high-speed camera is filtered to observe only IR reflections. To track the hands' pose, we add an off-the-shelf 3D pose estimator (LMC, \cite{LMC}), allowing for roughly 10ms latency pose estimations using a stereo pair IR cameras. We chose this sensor due to its availability, and because it operates in the IR spectrum. Note that any 3D pose estimator with similar performance can be used. We position the projector and camera in a collocated setup, and the LMC closer to the scene from above. The elements are mounted onto a rig rotated at roughly 45 degrees for tabletop viewing. The camera is covered with a 850nm IR pass filter, and a hot mirror is placed between it and the projector. System calibration is discussed in \Cref{sec:calibration}.

In terms of software, we add two additional steps to the traditional track-render-project pipeline (see \Cref{sec:motion_to_photon}), to compensate for misalignments, and perceptually reduce temporal artifacts. We first add a 2D deformation step to every rendered frame (\Cref{sec:mls_deformation}). This compensates for low-frequency tracking errors. We then introduce \textit{Perceptual Boundary Reduction} (\Cref{sec:pbr}), a just-in-time effect filling gaps between projected and observed hands, which constitute one of the most distinct visual artifacts, thus reducing perceived latency.

\subsection{Baseline Pipeline}
\label{sec:baseline}
\subsubsection{Calibration}
\label{sec:calibration}
\ourmethod~ requires three separate calibration steps: camera calibration, camera-projector coaxial calibration, and LMC-camera calibration.

For camera calibration, we use an off-the-shelf calibration procedure \cite{opencv_library}.

For coaxial calibration, the projector is fixed to the rig, while the camera is connected through a rotating mount. The camera is placed coaxially with the projector using the hot mirror to align the optical paths. Note the projector exhibits a very low principle point in screen space, so the camera must be tilted significantly. In a perfect coaxial setup a simple homography can describe the transformation between camera to projector space. In practical setups though, the optical elements are not perfect, and a homography only allows for a certain working volume where the two screen spaces align. We found a manual 4-corner point calibration according to a re-projected image of the scene more satisfactory than semi-automated methods \cite{coaxial}.

Lastly, to calibrate the LMC to the rest of the setup, we found tracking a small retro-reflector to be the most accurate and efficient method. The retro-reflector is glued to the tip of the index finger, and can easily be detected by both the LMC and camera in various locations. Its location is fed to a pose computation problem (SolvePNP, OpenCV \cite{opencv_library}) using a RANSAC approach. This allows us to obtain the camera extrinsics in LMC coordinate space. This calibration error reached a maximum of 5 mm in the peripheral of the LMC.

\subsubsection{Motion to Photon}
\label{sec:motion_to_photon}

For every projected frame, an LMC frame consisting of the individual bone-to-world transformations of each finger joint (referred collectively as the 3D pose) is computed by extrapolating the latest LMC frame using the LMC API. The desired timestamp of extrapolation is manually tuned to compensate for the full system latency. Hence, the extrapolated LMC frame roughly describes the state of the hand for the time the projected image will be displayed on the hand. Following this, the 3D pose is uploaded to the GPU and is used to linear blend skin a premade hand mesh model, with bones corresponding to the LMC bones. 
 
After skinning, any effect can be applied to the mesh (e.g. texturing, relighting, shading, baking, etc.) as long as the associated shader can run in reasonable time, which is not usually an issue as the mesh does not contain too many vertices ($2045$ vertices) and this step is not considered a bottleneck (see supplementary for a detailed profiling).

The \textit{baseline} approach is to now project the rendered result onto the scene. However, this yields unsatisfactory results due to misalignments. The next sections describe steps to improve the quality of projection significantly, using information from camera frames. Camera frames arrive every $\sim1.85ms$ (corresponding to the maximum 525 FPS of the camera). Frames are immediately uploaded to the GPU asynchronously, and used to augment every projected frame. We note that the max throughput of the projector (1000 FPS) is not being fully utilized in our system.

\subsection{Moving Least Squares Deformation}

\begin{figure}[h]
  \centering
  \includegraphics[width=1\hsize]{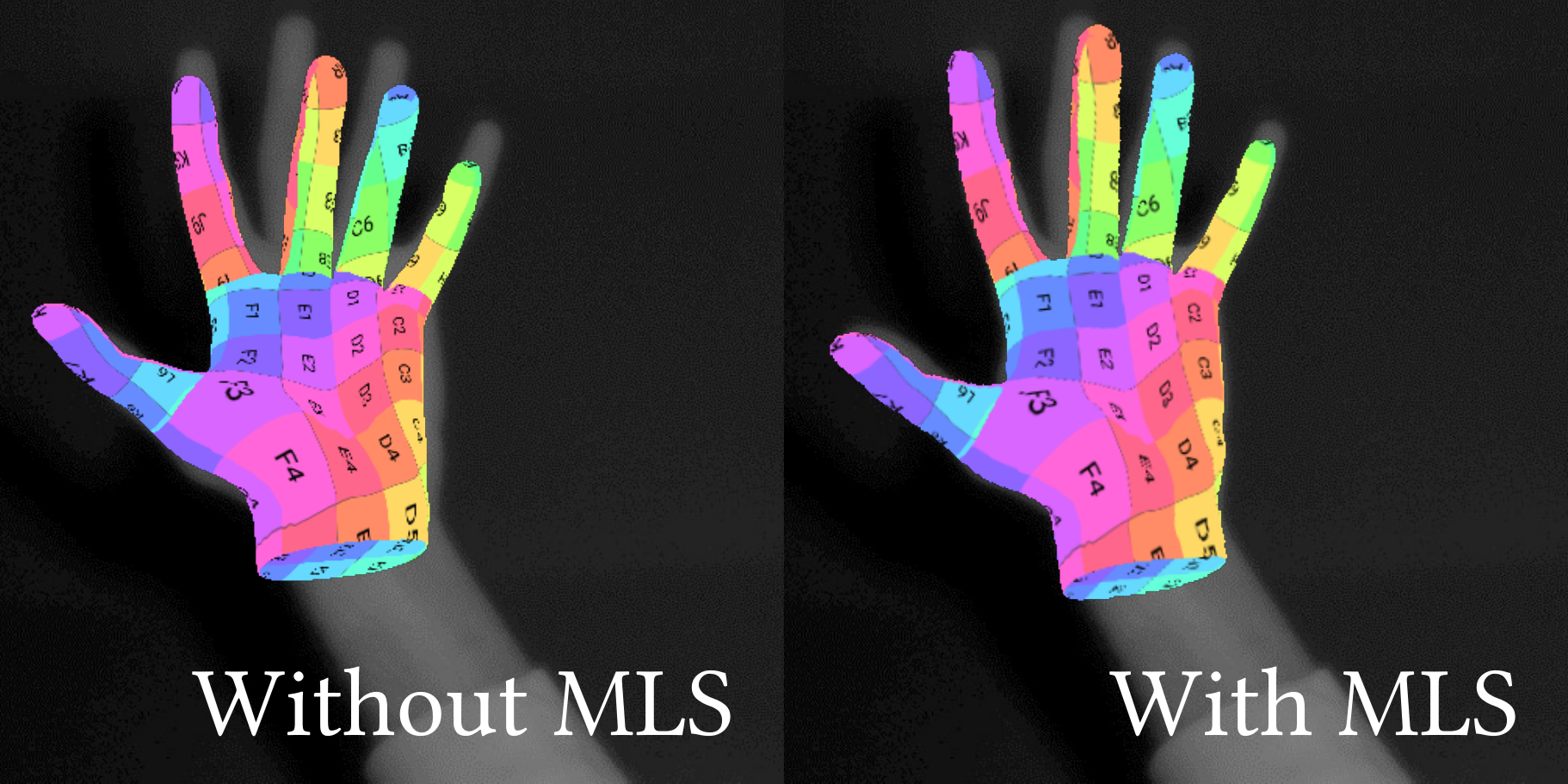}
  \caption{MLS deformation. Left: due to accumulated errors, the rendered hand generated from the 3D pose and the camera frame exhibit alignment errors. Right: we reduce the bias by applying a deformation grid in screen space, computed from a set of source points $Q$ (LMC Joints) and their target $P$ (2D landmarks).}
  \label{fig:mls_deformation}
  \Description{bias correction.}
\end{figure}

\label{sec:mls_deformation}
Using the LMC frames (or any other 3D pose estimator) directly to reconstruct a mesh, render it, and project it onto the hand potentially involves several unaccounted for errors: internal tracking errors from the estimator itself, calibration errors (between the LMC and camera), shape mismatch errors (between the 3D reconstructed shape and the actual hand, if a pre-defined static model is used), and errors associated with the skinning model (e.g. linear blend skinning). These errors, despite potentially being small individually, are very noticeable when aggregated (\Cref{fig:mls_deformation}, left), even with little to no movement. Our crucial observation is that we can alleviate most of this bias by employing a relatively slow but accurate correction step in screen space for slow movements or a completely static hand, and deal with more sudden and quick motion types in a separate manner which will be discussed in the next section. The choice of correcting in screen space is deliberate, as monocular image landmark estimators are much more accurate when they do not require to output depth and do not suffer from scale ambiguity. To this end, we leverage a 2D landmark detector \cite{mediapipe} based on a deep learning solution, and a uniform deformation grid used to deform the render. The purpose of the grid, is to deform the biased-render to fit the hand shape. We leverage a Moving Least Squares \cite{mls} discrete deformation grid, as its construction is extremely efficient, and only requires a set of source points $P$ and their target $Q$ as input. 

At any given time point $T$ we can project the location of the extrapolated LMC joints to screen space to obtain $P_{t=T}$. Note the joints are always consistent with the LMC bones, which are used for rendering. To obtain $Q$, we detect 2D landmarks of the hand from the current camera frame. In the ideal case, if the detector was instant, $Q_{t=T}$ would be obtained and used together with $P_{t=T}$ to compute the deformation grid. However, the detector is both compute-heavy (unable to run every frame, but rather only when the previous detection finished) and takes a relatively long time to run ($t_L\widetilde{=}16ms$). In the best case, the detector was launched at exactly $T-t_L$ and the landmarks used at time $T$ are in the past $Q_{t=T-t_L}$. Moreover, $Q_{t=T-t_L}$ will be used for all $T \leq t \leq T+t_L$ (until the next detector results are ready). Unfortunately, directly using $P_{t=T}$ and any $Q_{t<T}$ to solve the grid will result in lag, and using the same $Q_{t=T-t_L}$ for $P_{t>T}$ will result in jitter due to infrequent updates of the grid. Hence some form of temporal filtering is required. Since designing a temporal filter suitable for our setup is a complicated task, a full discrete simulation of our system was developed to determine the best solution. Our simulation provided several important insights which we discuss in \Cref{sec:temporal_filtering}. We found that the best and simplest solution is to propagate $Q_{T-t_L}$ forward in time using $\widehat{Q_{T}} = Q_{T-t_L} + P_{T} - P_{T-t_L}$ where $\widehat{Q_{T}}$ is our estimate of $Q_{T}$.

In practice, the detector is launched in a separate thread (when the previous detection is finished, due to compute limits), and provides us with $Q_{T-t_L}$. For each render loop iteration, we solve the grid using $P_T$ and $\widehat{Q_{T}}$ and interpolate the resulting deformation over the entire rendered image. This correction works especially well for the finger areas, which tend to be harder to track and are thin, making them difficult to reconstruct accurately. Results from this correction step can be seen in \Cref{fig:mls_deformation}. See supplementary for a more graphical illustration of using $P$s and $Q$s from different time stamps.

\subsection{Perceptual Boundary Reduction}
\label{sec:pbr}

\begin{figure}[h]
  \centering
  \includegraphics[width=1\hsize]{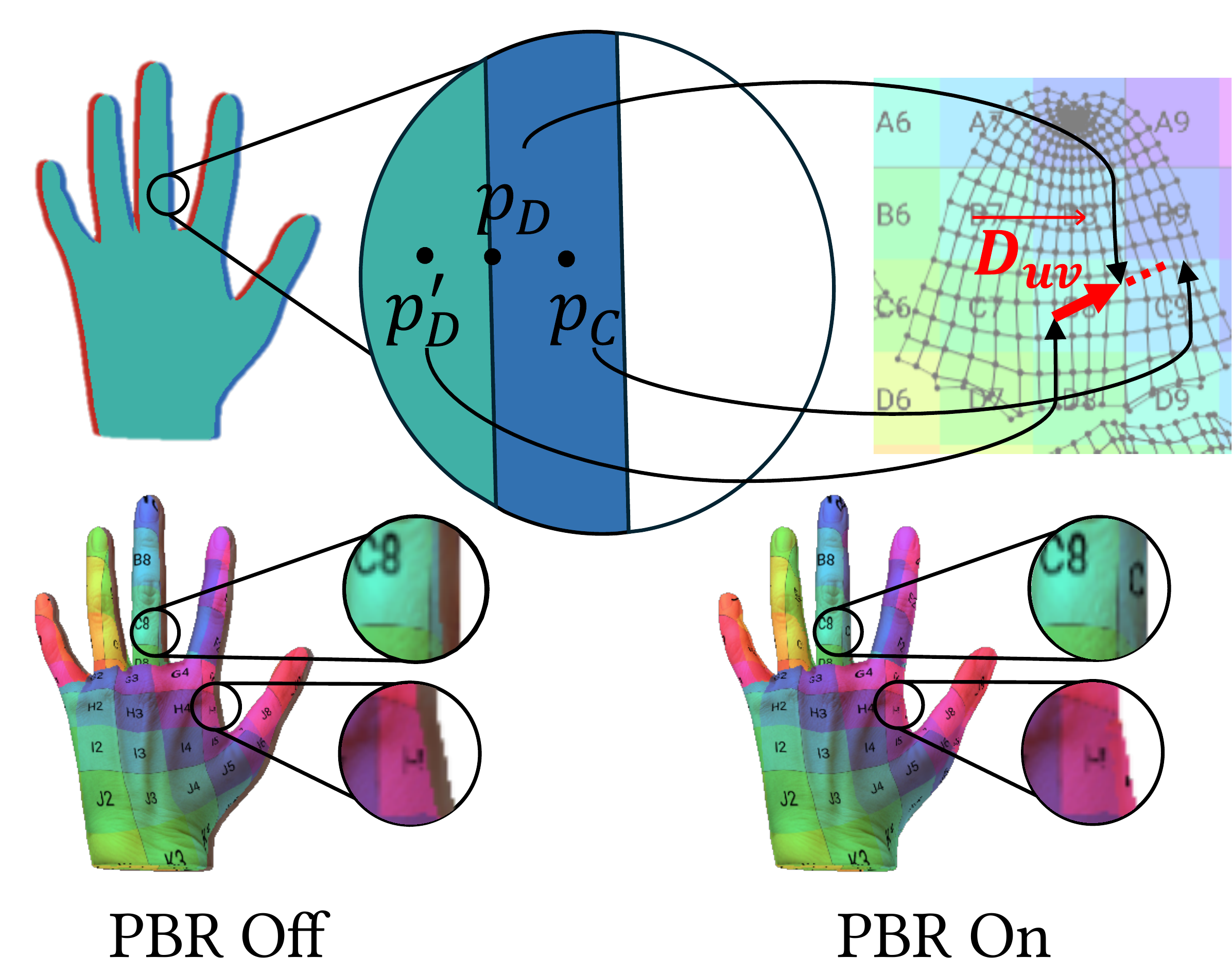}
  \caption{Illustration of the PBR step while the hand is translating to the right. Top: given the up-to-date {\color{regc}camera frame}, a  {\color{regb}rendered hand} and their {\color{regd}intersection}, we compute for each pixel $p_C \in \mathcolor{regc}{\Omega_C}$ the nearest neighbour $p_D \in \mathcolor{regd}{\Omega_D}$, and find a reflection about the neighbour $p'_D$. We then compute the UV coordinates of $p_C$ by extrapolating the UV coordinates of $p'_D$ and $p_D$. Bottom: comparing simulated frames when using versus not using the PBR step.}
  \label{fig:PBR}
  \Description{An image comparing the baseline method of projecting the rendered hand ontop of the camera frame, versus using the PBR step to fill in missing pixel information.}
  \vspace{-0.5cm}
\end{figure}

\begin{figure}[h]
  \centering
  \includegraphics[width=1\linewidth]{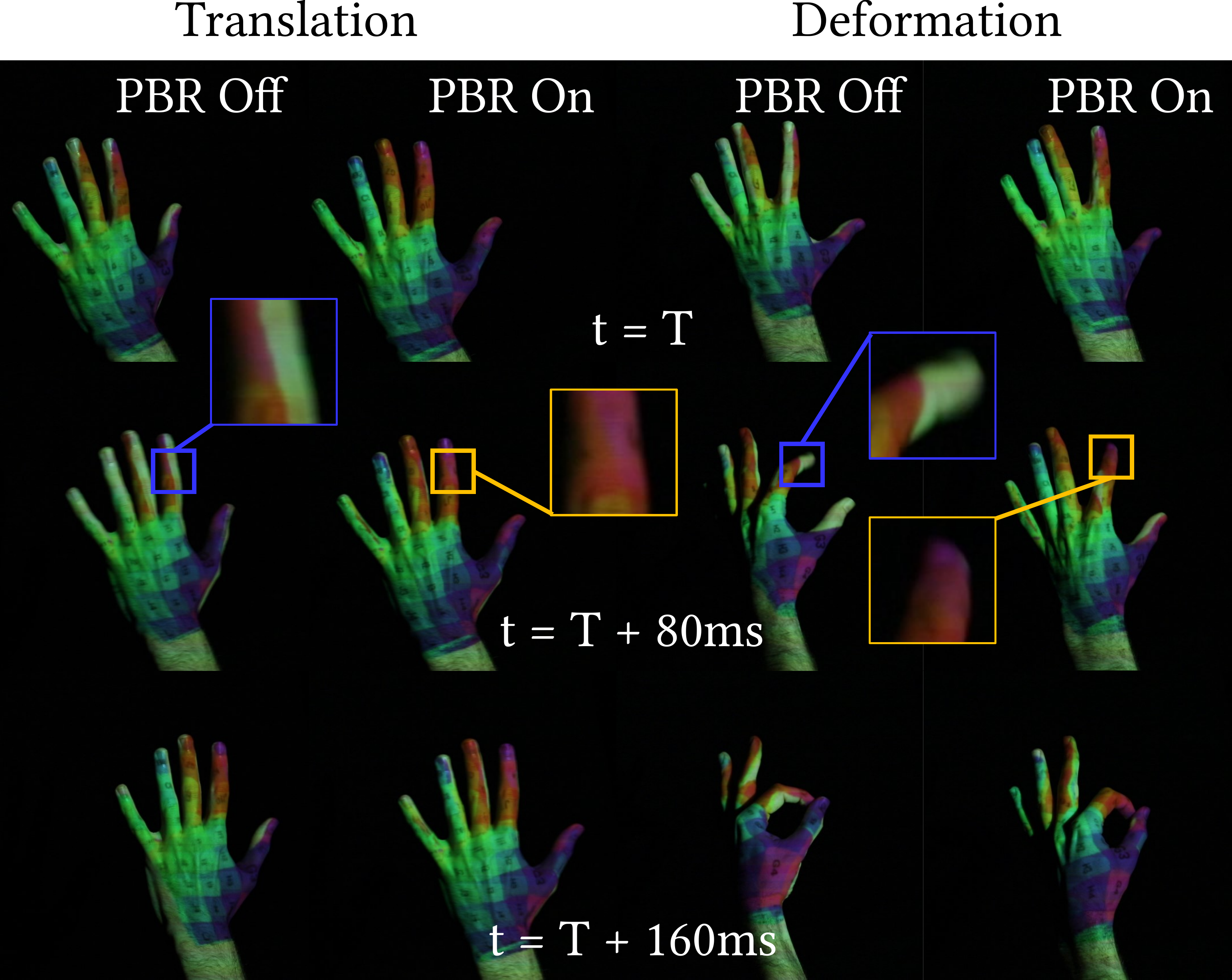}
  \caption{PBR During Rapid Motion. Left: the hand is translating from left to right. Right: the hand is deforming into the "OK" gesture. When using PBR, a visually plausible higher coverage of the hand is achieved both for static and moving hands. Note: the projected image in all cases is first deformed using the MLS step for better alignment.}
  \label{fig:PBR_motion}
  \Description{PBR during motion.}
\end{figure}

While the previous step improves alignment, there are still two existing issues. First, some visible areas of the hand, especially near the edges, are not covered by the projection. This is due to remaining shape mismatch errors. Second, the projection suffers from temporal misalignments during quick movements. This is because the most recent information processed at the time of projection originates from roughly $10ms$ in the past (coming from LMC frames). Extrapolation of the 3D pose helps reduce the effects of latency to some extent, but it relies on temporal filtering to approximate the current actual hand motion, and is susceptible to jitter and overshooting. While these particular issues are caused by the LMC latency, this phenomenon is general and unavoidable, as the current state of the art hand tracking is still not on par with the rigorous demands of projection mapping (see \Cref{sec:intro}). Our solution is to apply a Perceptual Boundary Reduction (PBR) step, addressing both issues. Our introduced step decreases the perceived notion of latency on the projected target and increases the overall surface available for projection, filling it with plausible information (see \Cref{fig:PBR} and \Cref{fig:PBR_motion}). We argue the hard seams formed by rendering edges on the hand are a strong perceptual cue for detecting misalignments. To this end, we disguise them using information from the surrounding area.

Consider the current camera frame while the hand is translating to the right. When superimposed on the rendered image (obtained from the baseline pipeline and applying the deformation grid), it partitions the image into four, not necessarily connected, distinct regions (see inset).

\setlength{\columnsep}{0pt}
\begin{wrapfigure}{r}{0.135\textwidth}
\includegraphics[width=0.13\textwidth]{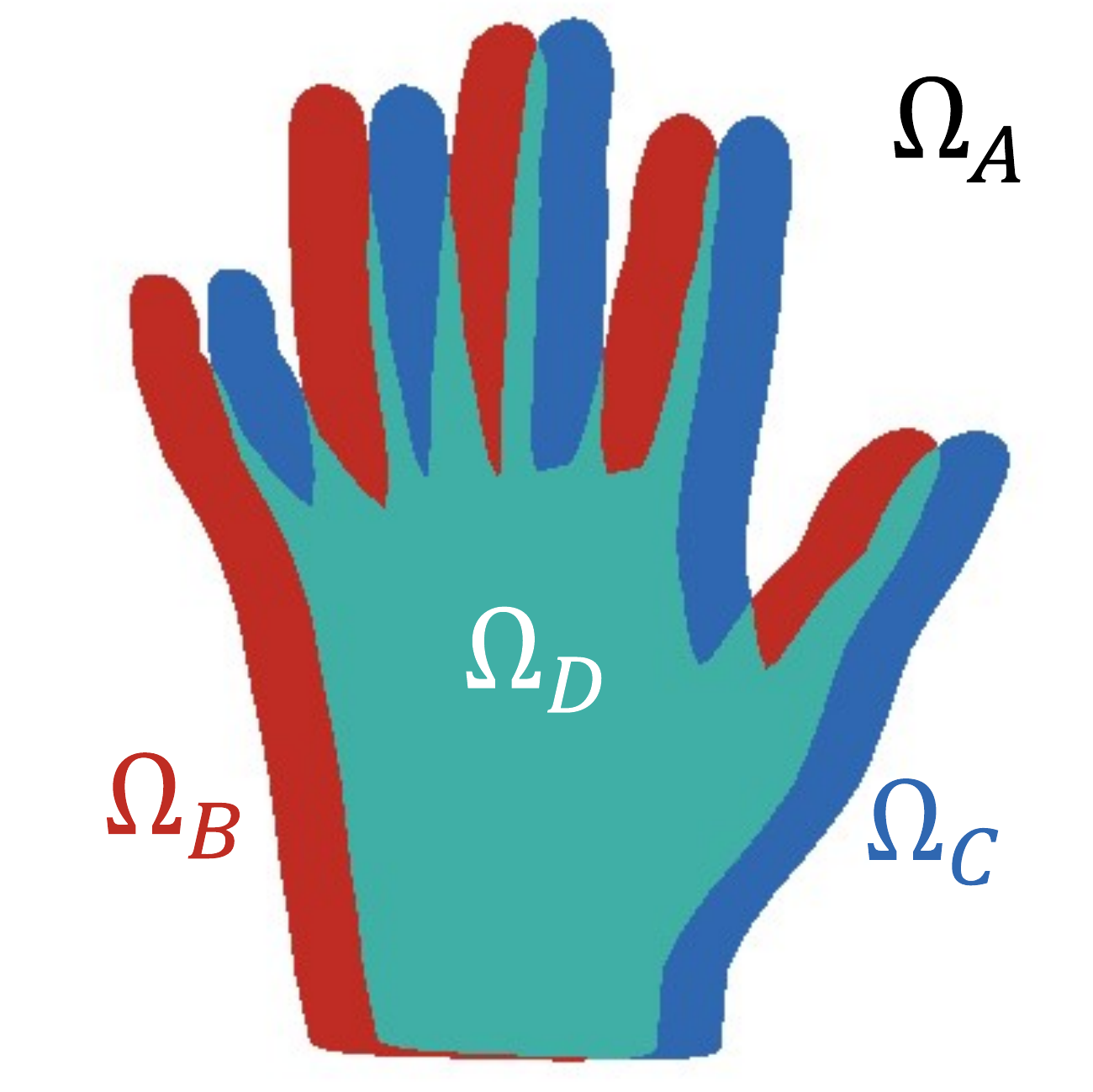}
\end{wrapfigure}
$\Omega_A$: Background pixels in the camera frame, which are also background pixels in the render, $\mathcolor{regb}{\Omega_B}$: background pixels in the camera frame, which are foreground pixels in the render, $\mathcolor{regc}{\Omega_C}$: pixels that depict a hand in the camera frame, but are background pixels in the render and $\mathcolor{regd}{\Omega_D}$: foreground pixels in both the render and the camera frame. Pixels in $\Omega_A$ clearly should stay background, while pixels in $\mathcolor{regb}{\Omega_B}$ should be masked in the final image as the camera frame is more up-to-date. Pixels in region $\mathcolor{regc}{\Omega_C}$, if left unattended, will form seams on the hand in the real scene, and pixels in region $\mathcolor{regd}{\Omega_D}$ should optimally be transformed to match the underlying geometry of the hand in the camera image. However, doing so robustly and quickly is a significant challenge as the hand does not contain strong surface features unlike other human body parts \cite{li2019survey}. Therefore, we opt to leave pixels in $\mathcolor{regd}{\Omega_D}$ unaltered, remove region $\mathcolor{regb}{\Omega_B}$, and address only region $\mathcolor{regc}{\Omega_C}$, reducing the perceived notion of latency by eliminating edges, and keeping everything else intact. Notice that the pixels in the region $\mathcolor{regc}{\Omega_C}$ will define all hard seams formed between the projected pattern and the hand ($\mathcolor{regd}{\Omega_D}\cap\mathcolor{regc}{\Omega_C}$). To fill the information for these pixels, we developed an approach relying on the nearest neighbours in $\mathcolor{regd}{\Omega_D}$ (that do contain rendered information): First, we jump flood \cite{jfa} all the pixels in this area to rapidly find the nearest available rendered pixel in $\mathcolor{regd}{\Omega_D}$. Then, instead of directly taking the nearest rendered pixel color value, we sample the mesh texture as if the current pixel was a natural extension of the nearest neighbour in the UV space (\Cref{fig:PBR}). To this end, we first project every pixel $p_C \in \mathcolor{regc}{\Omega_C}$ about its nearest neighbour $p_D \in \mathcolor{regd}{\Omega_D}$ in screen space, to obtain a new point $p'_D \in \mathcolor{regd}{\Omega_D}$, and retrieve the vector $\overrightarrow{D_{uv}} = f(p_D) - f(p'_D)$ where $f$ is the texture lookup operation (i.e. retrieving the $UV$ coordinates of a pixel). $\overrightarrow{D_{uv}}$ can be thought of as a linear approximation of the local gradient in UV space. We then obtain the new UV coordinate $f(p_C) = f(p_D) + \overrightarrow{D_{uv}}$ which samples the mesh texture as if $p_C$ is on the mesh by extrapolating the UV coordinates of $p_D$.

\section{Results}
\label{sec:results}
\subsection{Experimental Setup}
The computer controlling the system has a 4 core Intel Xeon(R) W-2223 CPU @ 3.6GHz, and an Nvidia GeForce RTX 2080 Ti GPU, with Windows 10 OS. The entire logic is written in C++ using a custom OpenGL engine. The projector is a three color channel Dynaflash \cite{dynaflash}, set to 1024x768 resolution, able of roughly 1000 FPS throughput and 3ms latencies, and connected to the PC using an optic cable and a custom driver. The camera is a Basler acA720-520um computer vision camera, connected through a USB port, and filtered using a 850nm cutoff-frequency low pass filter. The hot-mirror reflects >90\% of light in the range 750 - 1125nm, and transmits >85\% of light in the range 425 - 675nm. We use version 1 of the LMC and the Ultraleap Gemini (5th generation) platform for tracking. The IR illumination consists of a square array of 49 diffuse 850nm LEDs packaged together, originally purposed for night photography.

\subsection{Temporal Filtering}

\begin{figure}[h]
  \centering
  \includegraphics[width=1\linewidth]{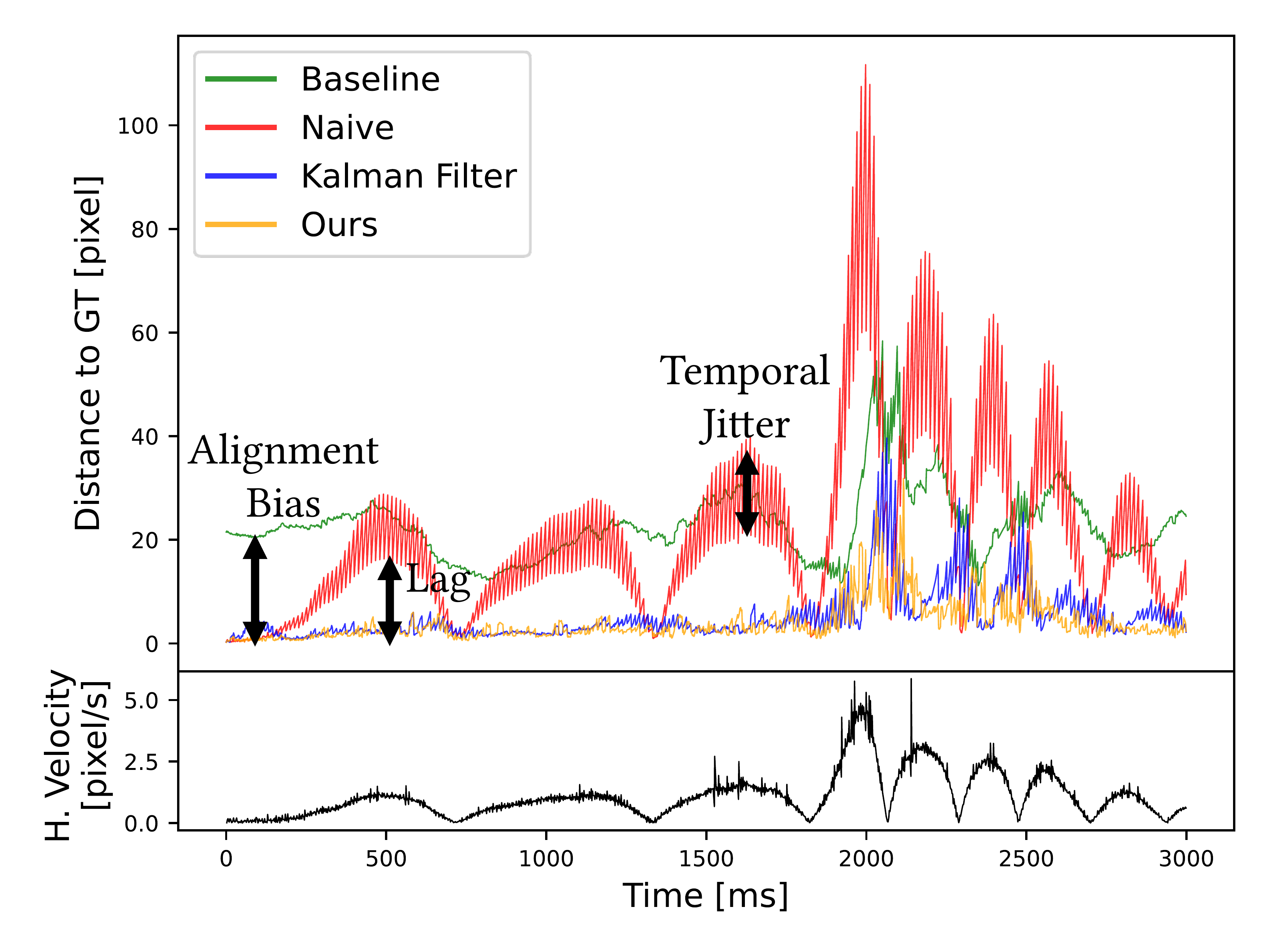}
  \caption{Discrete Simulation. Top: the error (in pixel units) of various temporal filtering approaches compared to the ground truth (the ideal case) averaged over all landmarks as a function of time. Bottom: absolute value of average landmark horizontal velocity. We ran our discrete simulation on prerecorded data of the left hand translating to the left and right for 3 seconds. The baseline solution where no MLS is used exhibits bias errors due to misalignment. The na\"ive method of not performing any temporal filtering on $Q$, reduces the average error but exhibits both lag and large amounts of jitter. Our approach is competitive with a Kalman Filter and simpler to compute, in part due to using information from previous $P$s rather than only previous $Q$s.}
  \label{fig:filters}
  \Description{A line plot showing the errors of different temporal filtering approaches for $Q$ as a function of time.}
\end{figure}

\label{sec:temporal_filtering}
We developed a discrete time simulation of our system to determine the best temporal filtering strategy, and investigate the behavior of the system as a whole. Results from this investigation is summarized in \Cref{fig:filters}. The simulation relies on prerecorded data of camera and LMC frames, obtained directly from our system. We ran the simulation for 5 seconds with different temporal filtering techniques, and compared them to the \textit{ideal} scenario where $Q_t$ can be obtained instantly every frame.

First, we verified that in the ideal scenario, the render is attached firmly and smoothly to the hand at all times, eliminating almost entirely any jitter or delay caused by extrapolating LMC frames using the LMC API. This is an important step confirming MLS deformation is indeed a feasible solution for removing bias. We computed a 2D B-spline curve approximation of all landmarks in the ideal scenario, and all reported distances are from this curve. The \textit{baseline} solution does not perform MLS deformation at all, and directly uses the render as the projection frame. This results in misalignments and large errors both visually and quantitatively in the simulation (see \Cref{sec:mls_deformation}). The \textit{na\"ive} solution is without temporal filtering, directly using $P_{t}$ and $Q_{t-x}$ to compute the deformation grid. Doing so results in both lag (as the target points are in the past) and jitter, as we only run the landmark detector when the previous detection is finished - so the grid is updated infrequently. For more elaborate filtering, we use a \textit{Kalman Filter} per point in the set $Q$ with a constant velocity motion model, which continuously gets updated by landmark measurements, and predicts their location into the future. We tuned the filter parameters to the best of our ability, yielding reasonable results both in simulation and in practice. However, using \textit{our} selected filter, we don't merely rely on the past information of $Q$, but incorporate information from how $P$ changed. This achieves competitive results with the Kalman Filter solution with much less compute. Our filter can also be seen as a Kalman Filter with velocity measurements from $P$ and setting a low measurement variance. Overall, the discrete simulation proved invaluable for quick experimentation and disqualification of filtering techniques instead of using the real system which is cumbersome and time-consuming.

\subsection{Just Noticeable Difference}

\begin{figure*}
    \centering
    \includegraphics[width=.99\textwidth]{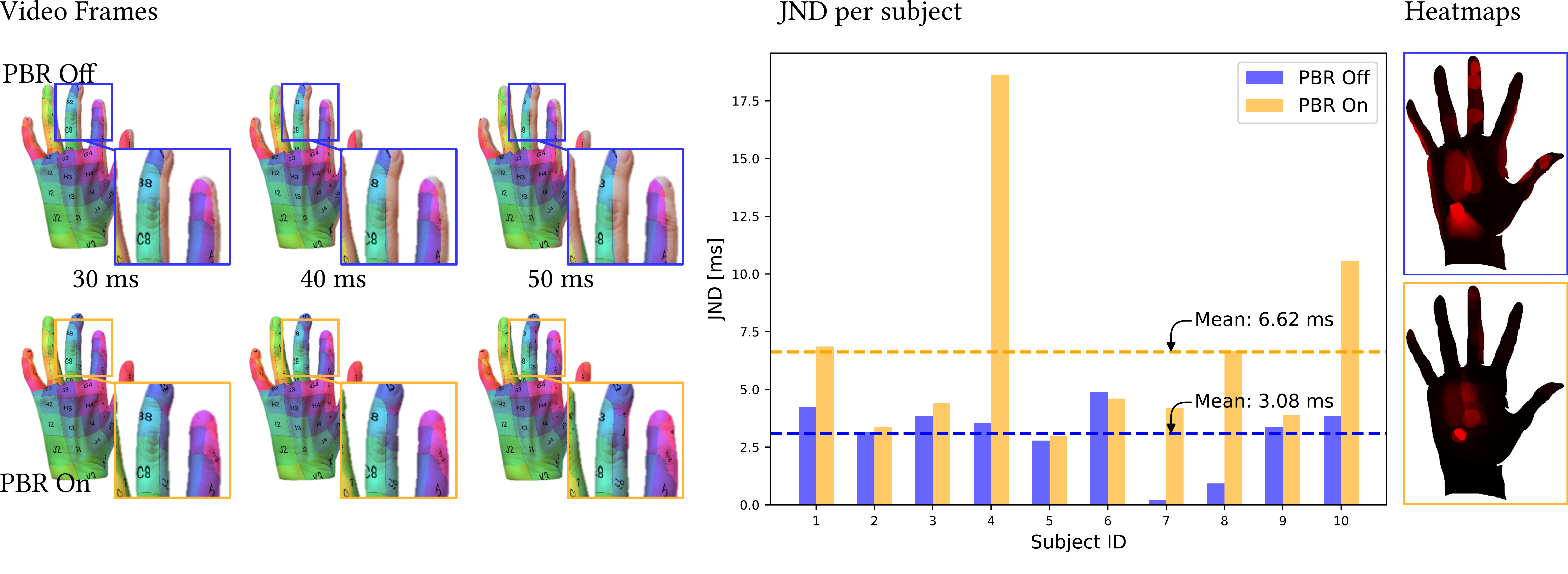}
    \caption{JND User Study Results. Left: an example frame from the videos shown during the user study with different simulated latencies. The PBR step fills the pixels with plausible information. Middle: Subjects exhibit significantly higher JND when the PBR step is used. Right: A heatmap of observed areas was computed from user markings. Red areas are locations where subjects marked they paid attention to. Our method successfully diverts subjects' attention from silhouette edges to the inner part of the hand, which is a less effective signal in determining latency as it relies on skin features.}
  \label{fig:jnd}
  \Description{The left image shows the JND user study results as a double bar plot (JND as a function of subject ID, baseline vs ours). The right shows two heatmaps of observed locations on the hand for the baseline vs our method.}
\end{figure*}

\label{sec:jnd}
Since our system is perceptually motivated and aims at reducing the perceived latency, we held a user study to estimate the just noticeable difference (JND, \cite{jnd}) metric over a simulation of projection mapping onto the hand with and without our PBR step. In a nutshell, this metric quantifies the minimum latency a subject can perceive in a given setup. The main problem with testing for JND directly with the live system, is that it exhibits non-negligible latencies, and is designed to project onto actual hands, rendering a controlled-repeatable experiment with short latencies impossible. Instead, we opted to use a prerecorded session of hand motion where the colorful patterns are synthetically projected onto it, and introduce artificial latency to the projection.

We followed the experimental protocol introduced by \citet{user_misalignment}, which is described in the following, together with a brief list of changes that were made to speed up the acquisition of data: subjects were exposed to a series (hereinafter a "session") of forced-binary choice trials, where each trial consists of a pair of videos projected onto a flat diffuse screen. The pair is one video with zero latency between the hand target and the projected image augmenting it, and one video with simulated latency. The pair was projected sequentially in random order, and both videos depict the same hand movements. Subjects were instructed to pick the video with more latency, where for each right decision we reduced the simulated latency in the next trial by a baseline step size, and for each wrong decision we increased it by threefold the baseline step. When users are correct in a trial, and wrong in the following trial (or vise-verse) this is called a "reversal". When this occurs, we half the baseline step, up to some minimum, and stop the session when ten such reversals occurred in total.

We made the following modifications to the original protocol that significantly reduced the experiment time per subject to a feasible one hour: Firstly, instead of prerecording videos before (or during) the experiment, we synthesized them on the fly using the same graphics engine used by the live system. Secondly, the starting simulated latency was set to 20ms instead of 40ms. Lastly, we did not use two interleaved sessions and average their result as originally proposed, but rather a single session. See supplementary for justification for these changes.

During the user study, participants were required to undergo 2 training sessions (with and without the PBR step), followed by 2 test sessions. The order of the sessions was determined by the Latin Square design to avoid bias. The prerecorded hand motion included all plausible motion types (translation, rotation, and deformation). After each session, subjects were also instructed to mark spots on the hand where they looked at the most to determine latency, by digitally painting red color on a simple illustration of the hand. We used this information to compute a heatmap of observed areas that were given attention to by subjects (see supplementary for full heatmap computation).

We recruited 10 participants aged $24 \pm 1.4 (avg \pm std)$, half of whom never saw dynamic projection mapping before. Subjects were seated at a distance from the screen such that the virtual hand performing the motion appeared roughly the same size as a human hand at arm's length. Results can be seen in \Cref{fig:jnd}. Average session time was $14.21 \pm 5.03 [min]$, subjects obtained a JND score of $3.08 \pm 1.38 [ms]$ when observing the baseline versus $6.62 \pm 4.54 [ms]$ using our method (a paired-t test indicate a significant difference with $t(9)=2.3$, $p<0.05.$), strongly supporting our claim that silhouette edges are an important signal for determining latency when projecting onto the hand, as opposed to using the skin features overlapping projection content inside of the hand to determine latency. We note one subject (ID=4) could not distinguish between perfect projection mapping (zero latency) and projection mapping using our method even with severe latencies ($>18ms$), but chose to include them in the analysis since this was a repeated result in both the training and test sessions. We also observed a clear trend in observation areas with and without using our method (\Cref{fig:jnd}, bottom) where our method successfully diverted the attention of the subjects to the inner parts of the hand, as silhouette edges between the projection and hand do not exist and cannot be relied on. Overall, the evidence suggest that using the PBR step provides significant slack for system latency, with little to no computational costs (for a detailed profiling of the system performance please see supplementary material).

\subsection{Guess the Character}

\begin{figure}[h]
    \centering
    \begin{subfigure}[b]{\columnwidth}
        \centering
        \includegraphics[width=\textwidth]{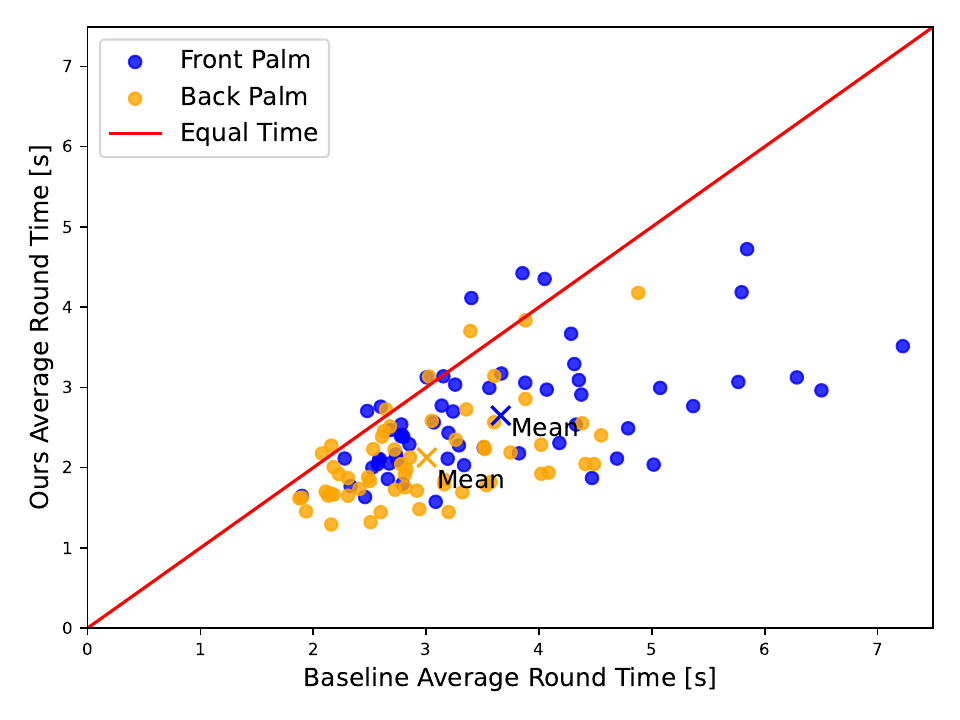}
        \label{fig:guess_char_scores}
    \end{subfigure}
    \vspace{-0.5cm}
    \begin{subfigure}[b]{\columnwidth}
        \centering
        \includegraphics[width=\textwidth]{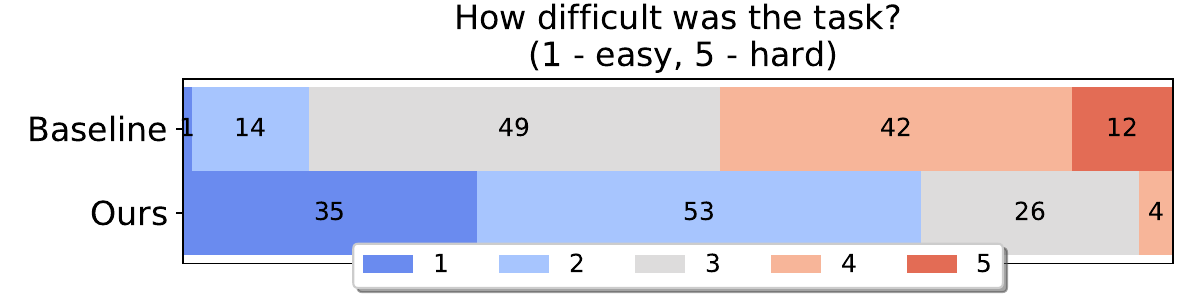}
        \label{fig:guess_char_a1}
    \end{subfigure}
  \caption{Guess the Character User Study Results. Top: The average round time using our method vs the baseline, for all subjects and sessions. In almost all sessions subjects were able to perform the task faster using our method both for the front and back facing palm configurations. Bottom: Subjects overwhelmingly found our method to ease the difficulty of the task, by answering the question "How difficult was the task?" immediately after each session.}
  \label{fig:guess_res}
  \Description{The pictures shows the Guess the Character user study results. The top side shows the average time it took for completing a round for all sessions and subjects, for our method (Y axis) vs without (X axis) in a scatter plot. The bottom side shows the result of asking the subjective question "How difficult was the task?" in a stacked bar plot.}
\end{figure}

\label{sec:guess_char}
We conducted a second user experiment, to asses whether users can perform tasks better when our correction steps are fully operational, versus the baseline (\Cref{sec:baseline}) --- projecting the skinned mesh constructed from the 3D pose estimator onto the hand.

We chose a task that demonstrates a dynamic environment, with special attention given to fine details on finger tips. Subjects were tasked with completing 24 sessions (12 with, and 12 without our correction steps), each consisting of 20 "rounds" to be completed as fast as possible. A round consists of projecting a random character onto the center of the hand, and the same character projected onto one of the finger tips. The other finger tips were also projected with other random characters, and the subject must bend the finger which has the same character as quickly as possible (see supplementary for visualization of the task). Additionally, the characters projected onto the finger tips randomly changed location in a small area every second to increase the task difficulty and improve robustness to misalignment. Subjects' accuracy and time to complete a session were recorded.

For this user study, 10 participants (age: $23.5 \pm 1.4$) were seated with the system above their shoulder, defining a working volume roughly in front of them and projecting onto their left hand. Between each session subjects were instructed to flip their hand, so that both the palm and backhand were used throughout the experiment. The baseline method and ours were interleaved, and the order was chosen using the Latin Square method. Subjects performed 4 training sessions before testing commenced (both methods, front and back facing hand). We also asked subjects to answer the question "How difficult was the task?" on a scale between 1 (easy) and 5 (hard) after each session, to determine subjective experience. 

Results can be seen in \Cref{fig:guess_res}. Subjects' round time using the baseline $3.33 \pm 1.0[s]$ decreased significantly using our method $2.39 \pm 0.71[s]$ (a paired-t test indicate a significant difference with $t(117) = 12$, $p<0.001$). Subjects overall felt performing the task was much easier using our method versus without.

\section{Applications}

\subsection{Space Shooter}

\begin{figure}[h]
  \centering
  \includegraphics[width=1\linewidth]{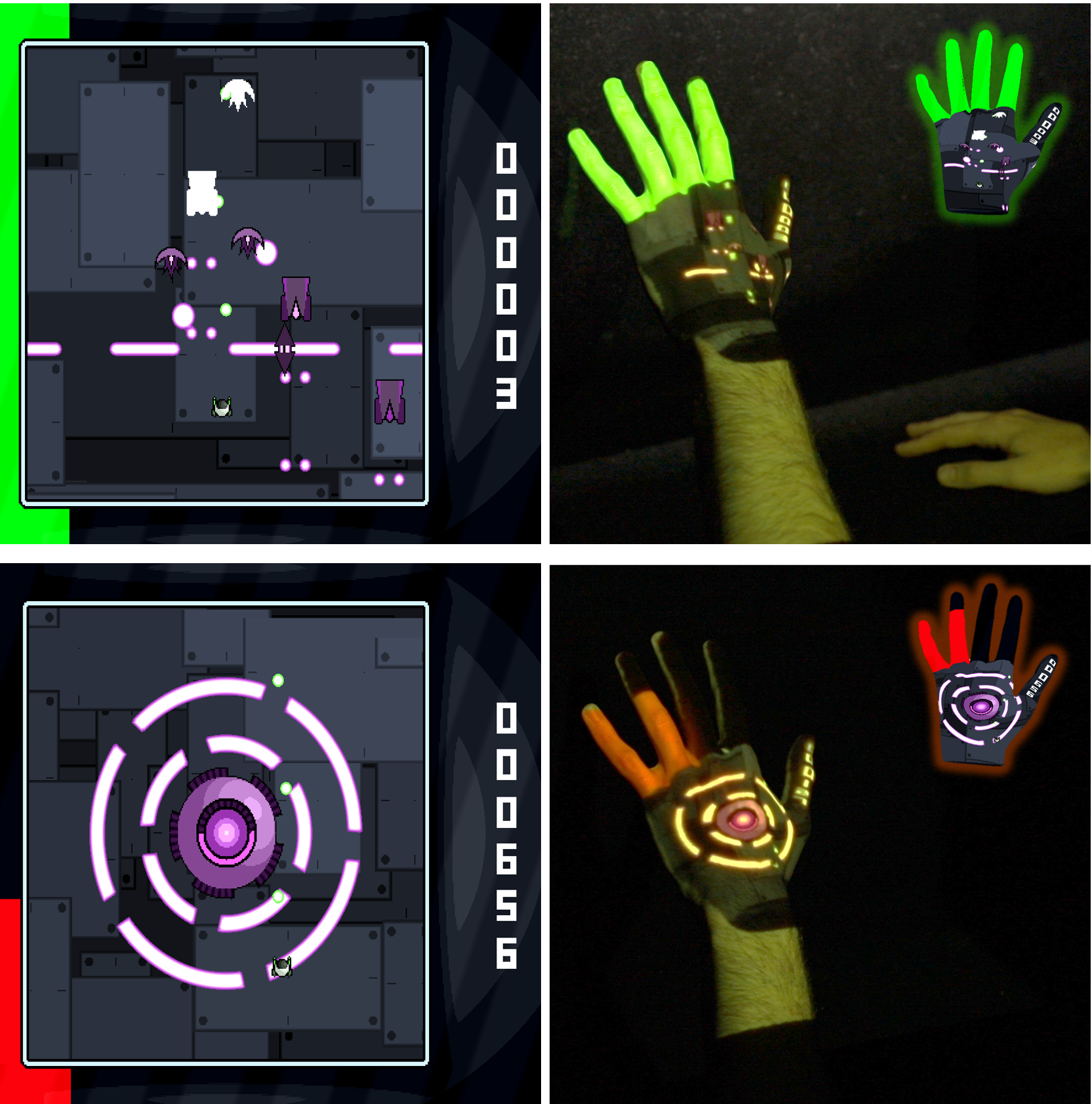}
  \caption{Space Shooter. Left: typical game frame rendering. Right: examples showing how a user interacts with the game, and the game frames wrapped around the hand mesh in inset.}
  \label{fig:spaceshooter}
  \Description{A figure showing on the left typical frame renderings. On the right the projections appears on the users hand while interacting with the game.}
\end{figure}

\label{sec:space_shooter}
We modified an open sourced space shooter game \cite{shadertoy} for the purpose of showing a simple gaming application with the system (see \Cref{fig:spaceshooter}). Players can move the space ship to the left or right by tilting their palm around the elbow-to-wrist axis, and shoot alien ships in front of the space ship. The score is displayed on the thumb, while the HP is visualized as bars displayed on the fingers, which decrease every time the spaceship is hit.

\subsection{Hand Projection Performance}

\begin{figure*}
  \centering
  \includegraphics[width=.99\textwidth]{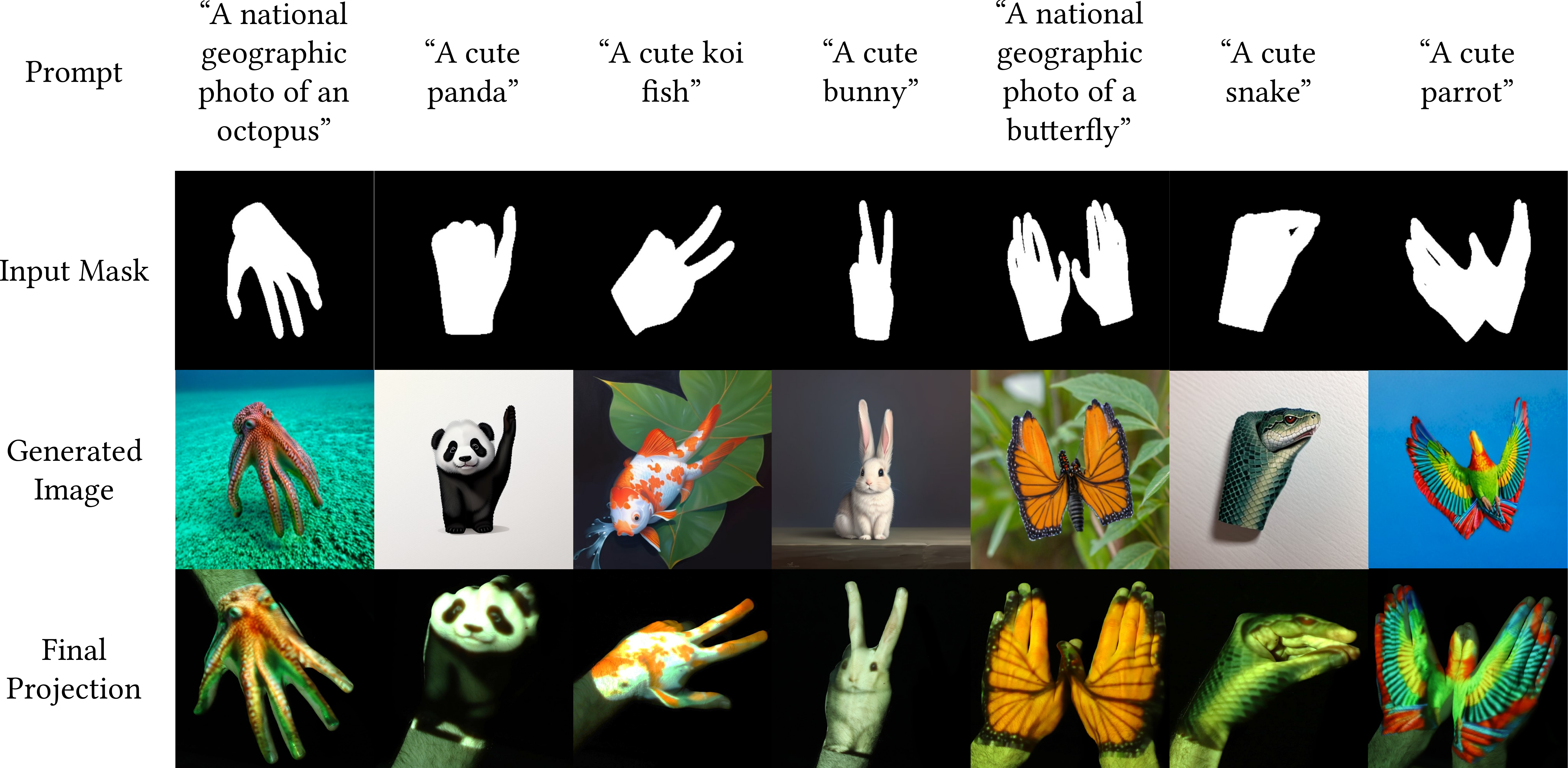}
  \caption{Projegraphy. A binary mask is created by skinning the hand mesh using the current LMC frame. The mask is then interpreted by ChatGPT4 which selects a plausible animal. A prompt is created using a predetermined stencil (e.g. "A cute ...", "A national geographic photo of ..."), and fed together with the rendered mask to ControlNet, which generates a plausible image. The image is baked onto the mesh and the result is projected.}
  \label{fig:animals}
  \Description{On the top, the image shows various automatically generated prompts used to generate images using ControlNet. The second row shows the binary mask used as a conditional input to ControlNet (which was also used to generate the prompt). The third row shows the generated images and the last row shows the final augmentation projected onto the hands.}
\end{figure*}

\label{sec:animals}
Inspired by \textit{Shadowgraphy}, the art of performing a story or show using images made by hand shadows, we developed a new concept art form for projecting live content onto hands depending on their shape and form. We name this \textit{Projegraphy}. Here, the hand or hands form some intricate shape, and the system generates suitable adaptive content for that shape. For example, the hand imitates some features of an animal, and the system projects relevant content without any additional user input. This can be used both for performances, and for social gaming ("guessing the animal"). Examples of Projegraphy for animals can be seen in \Cref{fig:teaser} and \Cref{fig:animals}. Note that the animals were not predetermined, and are generated on the fly depending on the hand shape and form.

To do this live during the performance, we first employ the multimodal model GPT-4 \cite{gpt4} to identify the animal that most resembles the current hand gesture by showing it a binary mask of the current hand silhouette. Then, ControlNet \cite{controlnet} conditioned on Canny edge images is utilized with the same mask as a control signal, to generate an image of the identified animal. We found that ControlNet performs better when the object occupies the majority of the frame. Hence, the mask undergoes preprocessing to ensure that the hand bounding box occupies a larger portion of the image. The generated image is baked onto the mesh using a projective texture and displayed using the real-time pipeline. This whole cycle takes between 2-8 seconds per animal and is launched on a separate thread, so the performer can animate the previously generated animal with no interference. See supplemental material for more details about the prompts and hyper parameters used to identify the animal and generate the images.

\section{Discussion}
This paper presented a fully functional short latency dynamic projection mapping system for human hands. We believe the solution to hold great potential for new applications in gaming, art and entertainment, and even medical rehabilitation. The key insight is that using information from a simple and rather coarse 3D sensor and correcting for its accuracy using cascaded 2D corrections is extremely efficient and only requires a small amount of computational effort. To the best of our knowledge, this is the first system to employ perceptual techniques in a dynamic projection mapping scenario, and we strongly believe this to be a good path forward for other dynamic scenarios such as rigid and semi-rigid bodies, where system induced latencies are even more pronounced relative to other sources of variance.

\subsection{Limitations and future work}

There are several limitations to our technique. First, it relies heavily on a reasonable coarse 3D estimation of the hand pose. When the estimation contains considerable error (e.g. wrong finger pose, large latencies) the MLS deformation and PBR steps are insufficient for recovery. Secondly, we employ a simple MLS deformation scheme to interpolate the sparse landmarks. This induces distortions and foldovers for extreme deformation, a problem that could be mitigated with better suited deformation schemes. In addition, the reliance on a premade 3D asset and animating it is not personalized enough, and inter-user hand variability (including shape, color and material properties of the skin) can reduce the quality of the projection. Accounting for such effects could increase the realism to a significant degree. Lastly, the PBR step relies on UV extrapolation, and is therefore sensitive to the mesh UV parametrization structure. If UV seams intersect the screen space vector $\overrightarrow{D} = p_D - p'_D$, then its UV space counterpart $\overrightarrow{D_{uv}}$ will fetch a color from a wrong region of the hand. In our implementation, we work around this by placing the seams in unnoticeable regions of the hand, and also hot-swapping the UV mapping with another when these regions become more visible (in our case, when the hand flips from front to back). 

\begin{acks}
We would like to thank Hiroki Kusuyama for helping with capturing the videos and pictures. \\
This work was partially supported by Len Blavatnik and the Blavatnik family foundation, and ISF (1337/22). \\
This work was partially supported by JSPS KAKENHI Grant Numbers JP20H05958.
\end{acks}


\bibliographystyle{ACM-Reference-Format}
\bibliography{egbib}

\end{document}